\documentclass[a4paper,11pt]{article}
\pdfoutput=1
\usepackage{jheppub}

\usepackage{graphicx}
\usepackage{placeins}
\usepackage{url}
\usepackage{multirow}
\usepackage{amssymb}
\usepackage{amsmath}
\usepackage{xspace}
\usepackage{booktabs}
\usepackage[symbol]{footmisc}

\usepackage{hyperref}
\usepackage{xurl}
\begin{document}

\title{Tools for Unbinned Unfolding}

\affiliation[a]{Department of Physics and Astronomy, University of California, Riverside, CA 92521, USA}
\affiliation[b]{Physics Division, Lawrence Berkeley National Laboratory, Berkeley, CA 94720, USA}
\affiliation[c]{Berkeley Institute for Data Science, University of California, Berkeley, CA 94720, USA}
\affiliation[d]{Department of Physics, University of California, Berkeley, CA 94720, USA}
\affiliation[e]{NERSC, Lawrence Berkeley National Laboratory, Berkeley, CA 94720, USA}

\author[a]{Ryan Milton,}
\author[e]{Vinicius Mikuni,}
\author[a]{Trevin Lee,}
\author[a]{Miguel Arratia,}
\author[b,d]{Tanvi Wamorkar,}
\author[b,c]{and Benjamin Nachman}

\abstract{
Machine learning has enabled differential cross section measurements that are not discretized.  Going beyond the traditional histogram-based paradigm, these unbinned unfolding methods are rapidly being integrated into experimental workflows.  In order to enable widespread adaptation and standardization, we develop methods, benchmarks, and software for unbinned unfolding.  For methodology, we demonstrate the utility of boosted decision trees for unfolding with a relatively small number of high-level features. This complements state-of-the-art deep learning models capable of unfolding the full phase space.  To benchmark unbinned unfolding methods, we develop an extension of existing dataset to include acceptance effects, a necessary challenge for real measurements.  Additionally, we directly compare binned and unbinned methods using discretized inputs for the latter in order to control for the binning itself.  Lastly, we have assembled two software packages for the OmniFold unbinned unfolding method that should serve as the starting point for any future analyses using this technique.  One package is based on the widely-used RooUnfold framework and the other is a standalone package available through the Python Package Index (PyPI).  


}

\maketitle

\section{Introduction}
\label{sec:introduction}

Differential cross sections are the currency of scientific exchange in particle, nuclear, and astroparticle physics.  These quantities connect the underlying theory to observations.  A critical step of any differential cross section measurement is the correction of detector distortions, also known as \textit{unfolding} or \textit{deconvolution}.  Traditional unfolding methods act on discretized data in the form of histograms.  These methods invert the folding equation  $m = R\,t$, where $t$ is the true histogram before detector distortions, $m$ is the observed histogram after detector distortions, and $R$ is the response matrix, which gives the probability of a measurement being in bin $i$ given the truth value is in bin $j$:
\begin{equation}
    R_{ij} = \text{Pr(measure in bin $i$ $|$ truth in bin $j$)}.
\end{equation}
The matrix $R$ is estimated using detector simulations. However, the solution $\hat{t}=R^{-1}m$ is not a good estimate because it amplifies statistical fluctuations, is not guaranteed to be non-negative definite, and may not exist if $R$ is not square. To circumvent this, various forms of regularized matrix inversion have been proposed over the years to perform unfolding.  See Refs.~\cite{Cowan:2002in,Blobel:2203257,doi:10.1002/9783527653416.ch6,Balasubramanian:2019itp} for reviews and Refs.~\cite{DAgostini:1994fjx,Hocker:1995kb,Schmitt:2012kp} for the most widely-used unfolding algorithms.

While histograms are a useful data representation for a specific analysis, they are fundamentally limited in scope. For example, bin boundaries cannot be changed after the measurement is published, thus inhibiting comparisons between experiments. Most importantly, while the data are intrinsically high-dimensional, published results are low-dimensional. This means that the histograms must be designed for a specific analysis, integrating over the correlations that are thought to be irrelevant for the particular case. For measurements of other observables, such as those motivated by theoretical insight originating long after the initial research, the analysis may have to be repeated from scratch.

This motivates machine learning (ML) based unbinned unfolding methods~\cite{Arratia:2021otl,Butter:2022rso2,Huetsch:2024quz}. Such techniques are based on discriminative~\cite{Andreassen:2019cjw,Andreassen:2021zzk,Pan:2024rfh} or generative models, 
either for distribution mapping~\cite{Datta:2018mwd,Howard:2021pos,Diefenbacher:2023wec,Butter:2023ira}
or for probabilistic, conditional generation~\cite{Bellagente:2019uyp,Bellagente:2020piv,Vandegar:2020yvw,Alanazi:2020jod,Backes:2022vmn,Leigh:2022lpn,Ackerschott:2023nax,Shmakov:2023kjj,Shmakov:2024xxx}.  These approaches do not require binning and can readily integrate many features to extend the phase space of the measurement and to improve the precision by accounting for any relevant correlations with the detector response.  In the last couple of years, there have been a number of ML-based unfolding measurements performed by various experiments, including H1~\cite{H1:2021wkz,H1prelim-22-031,H1:2023fzk,H1prelim-21-031}, LHCb~\cite{LHCb:2022rky}, ATLAS~\cite{ATLAS:2024xxl,ATLAS:2025qtv}, CMS~\cite{CMS-PAS-SMP-23-008}, and STAR~\cite{Song:2023sxb,Pani:2024mgy}.

While there has been some discussion of standard tools for presenting the results of ML-based differential cross section measurements~\cite{Arratia:2021otl}, there are currently no standard tools for doing the measurements themselves.  For classical unfolding, RooUnfold~\cite{Adye:2011gm,Brenner:2019lmf} is the community standard, used by a majority of all measurements in recent years across experiments.  The goal of this paper is to describe easy to use software for unbinned unfolding.  We have prepared two sets of tools.  In order to benefit from the extensive user base of RooUnfold, we have prepared a RooUnfold-inspired version of the OmniFold~\cite{Andreassen:2019cjw,Andreassen:2021zzk} ML-based unfolding algorithm. This currently exists in a fork of RooUnfold but it will be further developed for the inclusion of OmniFold into RooUnfold in the future. With this RooUnfold-inspired OmniFold, users of RooUnfold can now easily use ML by swapping out methods without changing any of the inputs or outputs of their analysis. To complement this, we have also prepared a standalone package in the Python Package Index (pip) that implements OmniFold.  The pip version is more light weight, since it does not rely on ROOT~\cite{Brun:1997pa}.  Best practices for the two approaches are discussed in more detail below.  We focus on OmniFold because it is the tool that has been used for all unbinned measurements so far.  The tools developed here are able to accommodate other methods, which may be added in the future.

This paper is organized as follows.  Section~\ref{sec:methodology} reviews unfolding methodology and the implementation of the machine learning methods. To the best of our knowledge, there has never been a direct comparison of OmniFold with classical techniques using the same inputs/outputs and also no comparison of OmniFold with decision trees versus neural networks. Numerical results of the latter comparisons in the context of Gaussian and collider physics examples are presented in Sec.~\ref{sec:results}, while comparisons using histograms as input to OmniFold and binned unfolding are shown in Appendix~\ref{sec:binned_unfolding}. The paper ends with conclusions and outlook in Sec.~\ref{sec:conclusion}. We additionally provide example usage of the unbinned unfolding methods in Appendix~\ref{sec:examples}.
\section{Methodology}
\label{sec:methodology}

As a classical baseline, we will compare unbinned unfolding with Lucy-Richardson Deconvolution~\cite{1974AJ.....79..745L,Richardson:72}, also known as Iterative Bayesian Unfolding (IBU)~\cite{DAgostini:1994fjx} in particle physics.  OmniFold generalizes IBU and so we start by reviewing IBU in Sec.~\ref{sec:ibu}. Section~\ref{sec:ML} then describes how pieces of IBU can be replaced with machine learning in order to process unbinned inputs and accommodate high-dimensional inputs/outputs.

\subsection{Iterative Unfolding}
\label{sec:ibu}

To begin, assume that there is no background to subtract and every event that passes the particle-level event selection passes the detector-level event selection (and vice versa).  Non-trivial acceptance effects are revisited at the end of this section.

For an initialized result $t^{(0)}_j=\Pr_0(\text{truth is in bin $j$})$, IBU proceeds as follows:

\begin{align}
  \nonumber
  t_j^{(n+1)} & = \sum_i \text{Pr}_{n}(\text{truth is $j$} \,|\, \text{measure $i$})
 \Pr(\text{measure $i$}) \\
  &= \sum_i \frac{R_{ij} t_j^{(n)}}{\sum_k R_{ik} t_k^{(n)}} \times m_i,
  \label{eq:unfolding}
\end{align}
where $n$ is the iteration number.  The result is $\hat{t}_\text{IBU}=t^{(N)}$ for some $N>0$.  As $N\rightarrow\infty$, the unfolded result converges to the maximum likelihood estimate~\cite{shepp1982maximum}.  In practice, the algorithm is truncated after a few iterations in order to regularize the result.

For what comes next, it is useful to rewrite the usual IBU presentation of Eq.~\ref{eq:unfolding} into two steps.  First, 

\begin{align}
\label{eq:step1IBU}
    \omega^{(n)}_i&=\frac{m_i}{\tilde{m}_i^{(n)}}\equiv\frac{m_i}{\sum_{k}R_{ik}t_k^{(n)}}\,,
\end{align}
where $\tilde{m}^{(n)}$ is the detector-level spectrum after the $n^\text{th}$ iteration.  The initial $\tilde{m}^{(0)}$ is the starting detector-level simulation.  The quantity $\omega$ are the weights required for the simulation to match the data.  In other words, when the weights $\omega^{(n)}$ are applied to $\tilde{m}^{(n)}$, it will match $m$.  The second step is then:

\begin{align}
\label{eq:step2IBU}
    \nu^{(n+1)}_j&=\frac{\tilde{t}^{(n)}_j}{t^{(0)}_j}\equiv\frac{\sum_i\Pr(\text{event in bin $j$ came from $i$})\,t_j^{(n)}\,w_i^{(n)}}{t^{(0)}_j}=\frac{\sum_i R_{ij}\,t_j^{(n)}\,w_i^{(n)}}{t_j^{(0)}}\,,
\end{align}

where $\tilde{t}^{(n)}$ is the particle-level spectrum induced by weighting events based on their detector-level weights $\omega_i^{(n)}$.  The final answer after $N$ iterations is then the starting simulation $t^{(0)}$ weighted by the final $\nu^{(N)}$, which is exactly the same as Eq.~\ref{eq:unfolding}.  Breaking IBU into these two steps looks more complicated, but makes manifest that there are two density ratios being approximated, one at detector-level for $\omega$ and one at particle-level for $\nu$.  OmniFold achieves unbinned unfolding by replacing $\omega$ and $\nu$ with machine learning tools, as described in the next section. 

Before proceeding, here is a brief note about background processes and acceptance effects.  Backgrounds are usually subtracted bin-by-bin from $m$ before the iterative process above begins.  Similarly, $m$ is modified bin-by-bin to account for events that pass the detector-level selection, but not the particle-level selection.  Finally, $t^{(N)}$ is scaled bin-by-bin to account for events that pass the particle-level selection, but not the detector-level selection.  These corrections ensure that the unfolding procedure compensates for experimental selection effects and represents the true underlying physical distribution.

\subsection{Unbinned Unfolding}
\label{sec:ML}

OmniFold is an iterative two-step procedure.  With continuous spectra, the goal is to estimate probability densities and not discretized histogram versions.  In the language of the previous section, $m\mapsto p_\text{data}(x)$, $\tilde{m}^{(0)}\mapsto p_\text{det.}(x)$, and $t^{(0)}\mapsto p_\text{part.}(z)$, where $x$ are the detector-level (det.) observables, $z$ are the particle-level (part.) observables.  The analog of Eqs.~\ref{eq:step1IBU} and~\ref{eq:step2IBU} in the continuous case are 

\begin{align}
\label{eq:step1OF}
    \omega^{(n)}(x)&=\frac{p_\text{data}(x)}{\tilde{p}^{(n)}_\text{det.}(x)}\equiv\frac{p_\text{data}(x)}{\int \text{d}z \, p_\text{det.}(x)\,\nu^{(n)}(z)}\,,
\end{align}

and

\begin{align}
\label{eq:step2OF}
    \nu^{(n+1)}(z)=\frac{\tilde{p}^{(n)}_\text{part.}(z)}{p_\text{part.}(z)}=\frac{\int \text{d}x \, p_\text{part.}(z)\,\omega^{(n)}(x)}{p_\text{part.}(z)}\,.
\end{align}
In practice, we only have samples from these densities, both observationally and in simulation.  The density ratios $\omega^{(n)}$ and $\nu^{(n+1)}$ are estimated from finite samples using machine learning classifiers.  As such, they are naturally unbinned and can readily be extended to many dimensions.  The $\omega$ density ratio is estimated by training a classifier to distinguish data from detector-level simulation where each event is weighted by the $\nu$ from the previous iteration.  The integral in the denominator of Eq.~\ref{eq:step1OF} reflects the fact that the weights are determined by $z$, but the classifier acts on $x$, so we need to use the fact that in simulation, the particle-level and detector-level events are paired.  Similarly, the density ratio $\nu$ is estimated by training a classifier to distinguish the starting, particle-level simulation from the same simulation with each event weighted by the $\omega$ from the previous iteration.  The final result is a set of particle-level events sampled from $p_\text{part.}$, weighted by $\nu^{(N)}(z)$.

When $x$ and $z$ are one-dimensional, the algorithm is called \textsc{UniFold}; when they are multi-dimensional, the algorithm is referred to as \textsc{MultiFold}.  If all observables are simultaneously unfolded, the algorithm is labeled \textsc{OmniFold}.  When it is clear from context, the algorithm is simply referred to as \textsc{OmniFold}.

We have built two tools that implement OmniFold.  The first tool uses Boosted Decision Trees (BDTs) for the classifiers and is designed based on the RooUnfold framework. Including this version of OmniFold into RooUnfold is a future step. All current differential cross section measurements use \textit{tabular data}, i.e. a fixed set of high-level observables. It is well-known in the machine learning literature~\cite{grinsztajn2022treebasedmodelsoutperformdeep} and increasingly studied in particle physics~\cite{Finke:2023ltw,Freytsis:2023cjr} that BDTs outperform neural networks for tabular data.  They are also fast, stable, and have few hyperparameters.  Consequently, GPUs or other hardware accelerators are not required to train these models.  Due to their structure, no feature standardization is required, as is often necesary for neural networks.
%
%
RooUnfold is the most widely-used tool for unfolding and so by developing a version of OmniFold that is based on RooUnfold, we enable a diverse set of users to try out new methods with little overhead.  At the same time, we have also created a pip-installable tool that uses neural networks.  Neural networks are well-suited for continuous data and scale well to many dimensions.  In particular, state-of-the-art neural networks are uniquely suited for measurements of the full, variable-length, high-dimensional phase space.  Such measurements would simultaneously determine the spectra of all outgoing particles from a collision.  Furthermore, this tool is completely independent of ROOT, which makes it easier to use for a broader audience.  We believe that these two tools are complementary and the user should pick the best tool suited for their application.  As shown in Sec.~\ref{sec:results}, both tools have a similar performance on common tasks.

The rest of this section describes some technical details about the BDT and neural network implementations in our two software tools. Examples on how to run these tools in Python are given in Appendix~\ref{sec:examples}.

\subsubsection{RooUnfold-inspired BDTs}

Our implementation uses boosted decision trees from scikit-learn~\cite{scikit-learn}. We use GradientBoostingClassifiers with the default parameters -- a maximum depth of 3, 100 boosting stages, a learning rate of 0.1, and a cross-entropy loss. To prevent overfitting, early stopping can be enabled based on the model's performance on a validation dataset consisting of a fraction of the input data. The optimal parameters for early stopping are dependent on the dataset and should be determined by the user based on the model performance. Early stopping is not used in the numerical results of Sec.~\ref{sec:results} and Appendix~\ref{sec:binned_unfolding}.

Background subtraction and accounting for events that pass the detector-level selection, but not the particle-level selection are currently not implemented unbinned. 
It is often possible to reduce the contribution of events that pass the detector-level selection, but not the particle-level selection by having a more inclusive particle-level selection and then cropping the fiducial volume at the end to reduce the impact of events that are not well-constrained by data.  The converse is not advisable, so we have implemented unbinned corrections for events that pass the particle-level selection, but not the detector-level selection following Ref.~\cite{Andreassen:2021zzk}.  In particular, when assigning weights $\omega$ in the second step of OmniFold to events that do not pass the detector-level selection, we choose the average over events that do pass.  In other words, for an event $i$ with particle-level $z_i$ but no corresponding $x_i$, $\omega^{(n)}\equiv \langle \omega^{(n)}(x)|z=z_i\rangle$, where the average is computed over events that pass the detector-level selection.  To compute the average, we use a GradientBoostingRegressor with the same parameters as the GradientBoostingClassifiers.  While different approaches were not extensively studied, using the average weight seems to work well for BDTs compared to leaving the weight untouched from the previous iteration as is done for the neural networks described in the next section.

In addition to the unbinned studies presented in Sec.~\ref{sec:results}, we cross-check the BDTs with binned (i.e. discretized) inputs in Appendix~\ref{sec:binned_unfolding}. Similar to IBU, this uses the response matrix and a histogram of the experimental data as input. The BDT binned unfolding and IBU agree well, up to stochastic aspects of the BDT training. It was already known formally that these two should be the same, but we believe this is the first explicit comparison using the same inputs for both methods.

\subsubsection{Deep Neural Networks}

Our implementation uses \textsc{Keras}~\cite{chollet2015keras} and \textsc{TensorFlow}~\cite{tensorflow2015-whitepaper}, where there are pre-defined models as well as the option to input a custom model.  There is a built-in data loader and the iterative steps and network training are wrapped in easy-to-use classes.  One of the default models is a fully-connected feed-forward network that is appropriate for fixed-length inputs (e.g. tabular data).  The second model is based on the the Point-Edge Transformer~\cite{Mikuni:2024qsr} that supports point-cloud inputs (e.g. a variable-length list of four-vectors).  The details of the implementations can be found in the software, linked at the end of the paper. In the results shown in Sec.~\ref{sec:results}, we refer to the fully-connected feed-forward network as DNN and the Point-Edge Transformer as PET.

\section{Numerical Examples}
\label{sec:results}

\subsection{Gaussian Case}

We begin with a one-dimensional unfolding example, using spectra that follow those given in the RooUnfold tutorial~\cite{RooUnfold_Tutorial}.  In particular, the particle-level simulated events are sampled from a Breit-Wigner distribution with a mean of 0.3 and a width of 2.5 and the particle-level pseudodata events come from a Gaussian distribution with a mean of 0 and a standard deviation of 2. These particle-level spectra are smeared with a Gaussian of mean -2.5 and a standard deviation 0.2 to generate the detector-level events. To simulate efficiency, a cut is made in which events with detector-level values above the value $x_\text{eff}$ have their detector-level value removed. The efficiency value is given as:
\begin{equation}
    x_\text{eff}(x) = 0.3 + \frac{0.7}{20}(x + 10),
\end{equation}
where $x$ is the particle-level value.

This dataset uses 2 million simulated events and 500,000 pseudodata events. All of the events are given to IBU, while 1 million simulated events and 375,000 pseudodata events are used for training the ML methods and the remaining are used as test data. The efficiency cuts remove 39\% and 35\% of events from the detector-level simulation data and the detector-level pseudodata, respectively.

We construct a response matrix using the particle-level and detector-level simulated data. We also make a histogram containing the detector-level pseudodata. All the binning uses 40 bins linearly spaced between -10 and 10. These are used as input for the IBU unfolding.

For the unbinned unfolding, we store these data as lists and use a mask to account for the efficiency cuts.

\begin{figure}
    \centering
     \includegraphics[width=0.6\textwidth]{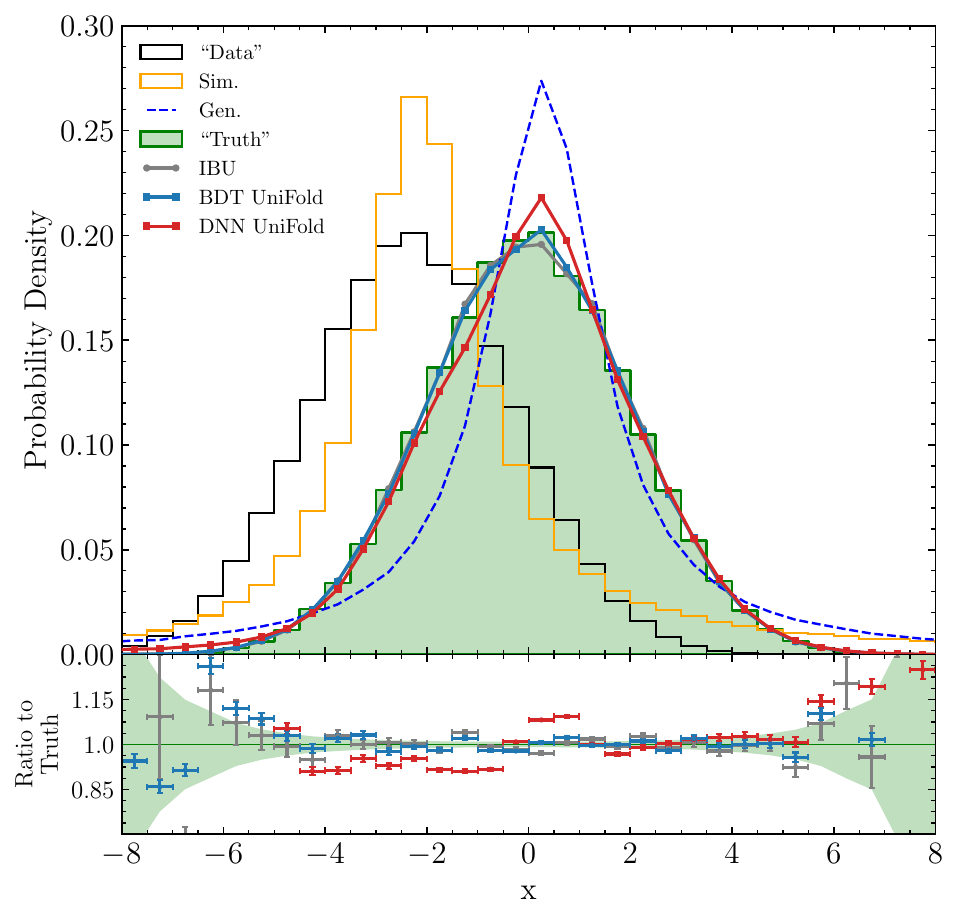}
     \label{fig:unfolded_gaussian}
     \caption{Comparison of unfolded Gaussians. Top panel: The distributions of the used datasets and the unfolded distributions using IBU, BDTs, and DNNs. Bottom panel: Ratio of the unfolded distribution to the truth data in each bin. The uncertainties in each bin are the quadrature sum of the weights for events in that bin (the weights are 1 for non-unfolded results), divided by the truth bin entries. The uncertainty band shows $1 \pm $ the truth uncertainty. }
\end{figure}

The results of the unfolding are shown in Fig.~\ref{fig:unfolded_gaussian}.  In line with previous papers on unbinned unfolding, the particle-level simulated data is called ``Gen." while the particle-level pseudodata spectrum is called ``Truth".  The detector-level samples corresponding to the Gen. and Truth are ``Sim." and ``Data", respectively.   Note that the UniFold results are binned for illustation purposes only -- the actual unfolded spectra are unbinned and can be e.g. rebinned on the fly.

All the unfolding methods show good agreement with the truth values in the majority of the distribution -- all within about 10\% of the truth. Towards the tails, the agreements with truth worsen due to the lack of data and low efficiencies in these regions.  Due to the stochastic nature of training, neural networks have an additional source of bias compared to other methods that can be mitigated with ensemlbing or other approaches.  No ensembling is performed here.  The BDT approach requires no ensembling and has the same or better accuracy as IBU across the full range.  While BDTs are not able to effectively handle full phase space unfolding (next section), they may be the best tool for unfolding high-level observables in the form of tabular data~\cite{grinsztajn2022treebasedmodelsoutperformdeep}.

\subsection{Jet Observables from Particle Colliders}
We extend our study to jets produced in proton-proton collisions at $\sqrt{s} = 14 ~\mathrm{TeV}$~\cite{Zenodo:EnergyFlow:ZJetsDelphes} -- the same dataset used in Ref.~\cite{Andreassen:2019cjw}. The  simulated data is generated with \textsc{Pythia 8.243} tune 26~\cite{Sjostrand:2014zea} and the pseudodata comes from the default tune of \textsc{Herwig 7.1.5}~\cite{Bellm:2017bvx}. To produce the detector-level events, we use the \textsc{Delphes 3.4.2}~\cite{deFavereau:2013fsa} fast simulation of the CMS detector. The particles from these events are clustered into jets using the anti-$k_T$ algorithm~\cite{Cacciari:2008gp} in \textsc{FastJet} 3.3.2~\cite{Cacciari:2011ma} with radius $R = 0.4$. These datasets contain the leading jets from each event and their constituent particles. We obtain approximately $1.6 \times 10^6$ events from each generator. Further details on the dataset are in Ref.~\cite{Andreassen:2019cjw}.

To investigate the unfolding performance, we consider six widely-used jet substructure observables: jet mass $m$, constituent multiplicity $M$, the $N$-subjettiness ratio $\tau_{21} = \tau_2^{(\beta = 1)}/\tau_1^{(\beta = 1)}$~\cite{Thaler:2010tr}, the jet width $w$, the groomed jet mass $\ln \rho = \ln m^2_\text{SD} / p_T^2$, and the momentum fraction $z_g$~\cite{Larkoski:2014wba}. The jet grooming is done with Soft Drop (SD) with $z_{cut} = 0.1$ and $\beta = 0$.

To illustrate the different ML methods, we perform \textsc{OmniFold} with the PET using all the reconstructed particles in each event, \textsc{MultiFold} with the DNN and BDT to simultaneously unfold all six observables, and \textsc{UniFold} with the BDT to unfold each observable individually. For training these, we use 1.2 million of the data. For the PET and DNN, 20\% of this is reserved as a validation set during training. The remaining $~400\mathrm{k}$ events are used to test each model's performance after training. For IBU, we use all 1.6 million events as input. For all methods, we also include a jet $p_T$ cut in which detector-level events that have $p_T < 150~\mathrm{GeV}$ are removed from their respective dataset, removing about 15\% of the events from both the simulated data and pseudodata.  This introduces acceptance effects that are absent from the original OmniFold paper~\cite{Andreassen:2019cjw} (and later studied with Gaussian examples in Ref.~\cite{Andreassen:2021zzk}).  While the $Z$ boson selection gives a nearly unbiased approach for selecting jets, the jet $p_T$ threshold mimics typical triggering requirements for more inclusive selections.

The neural networks are trained with Keras~\cite{chollet2015keras} and TensorFlow~\cite{tensorflow2015-whitepaper} using the Adam optimizer~\cite{Adam} over 50 epochs with a batch size of 256 and a learning rate of $10^{-4}$. All unfolding methods are compared using $n=5$ iterations, based on the original \textsc{OmniFold} paper~\cite{Andreassen:2019cjw}. None of these settings have been extensively optimized.  As a result, the numerical results that follow are meant to give a sense for the relative performance without being definitive.  The settings of each method should be optimized per application.

\begin{table}[h]
\centering
\footnotesize
\caption{Unfolding performance using triangular discriminator. We also show the metric for the detector-level pseudodata and particle-level simulation data. The best unfolding method for each observable is shown in bold.
}
\begin{tabular}{c c c c c c c}
     \hline
     \hline
     & \multicolumn{6}{c}{Observable} \\
     \cline{2-7}
     Method & $m$ & $M$ & $w$ & $\ln \rho$ & $\tau_{21}$ & $z_g$ \\
     \hline
     \textsc{OmniFold} PET & \textbf{0.79} & \textbf{0.42} & 0.81 & 0.66 & 1.97 & 0.74 \\
    \textsc{MultiFold} DNN & 2.38 & 0.46 & 0.35 & \textbf{0.20} & \textbf{0.63} & 0.47 \\
    \textsc{MultiFold} BDT & 2.64 & 0.53 & \textbf{0.31} & 0.22 & 0.92 & 0.74 \\
    \textsc{UniFold} BDT & 9.51 & 1.44 & 0.46 & 0.73 & 1.11 & \textbf{0.44} \\
    IBU & 9.74 & 1.46 & 0.48 & 0.89 & 1.15 & 0.45 \\
    \hline
    Data & 24.59 & 129.92 & 15.87 & 14.70 & 11.12 & 3.86 \\
    Generation & 3.46 & 14.94 & 22.39 & 18.97 & 20.70 & 3.99 \\
     \hline
     \hline
\end{tabular}
\label{table:metric}
\end{table}

\begin{figure}[h]
    \centering
    \begin{minipage}{0.5\textwidth}
        \centering
        \includegraphics[width=\linewidth]{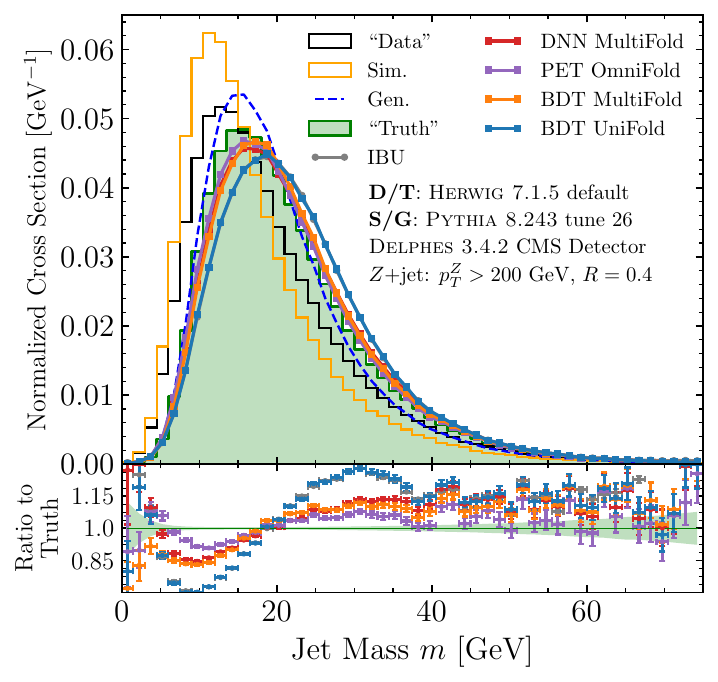}
    \end{minipage}\hfill
    \begin{minipage}{0.5\textwidth}
        \centering
        \includegraphics[width=\linewidth]{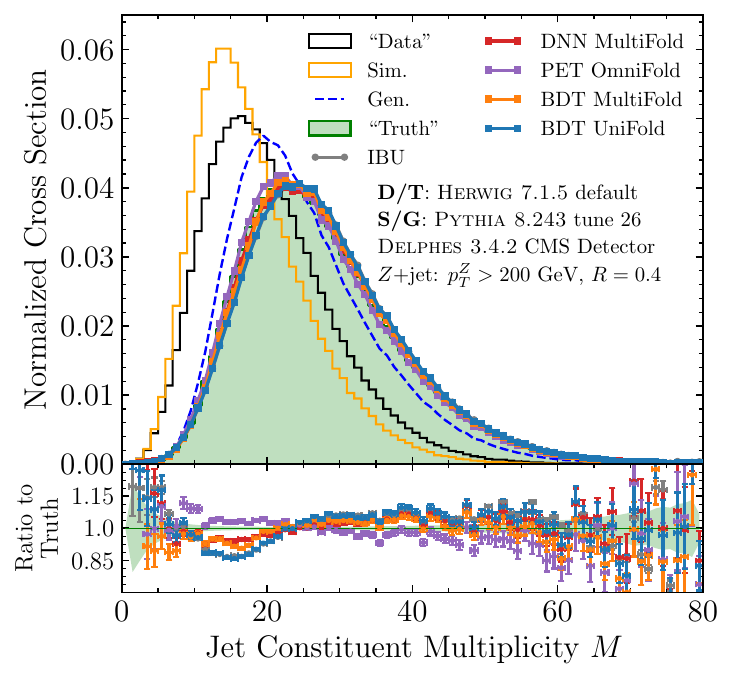}
    \end{minipage}

    \begin{minipage}{0.5\textwidth}
        \centering
        \includegraphics[width=\linewidth]{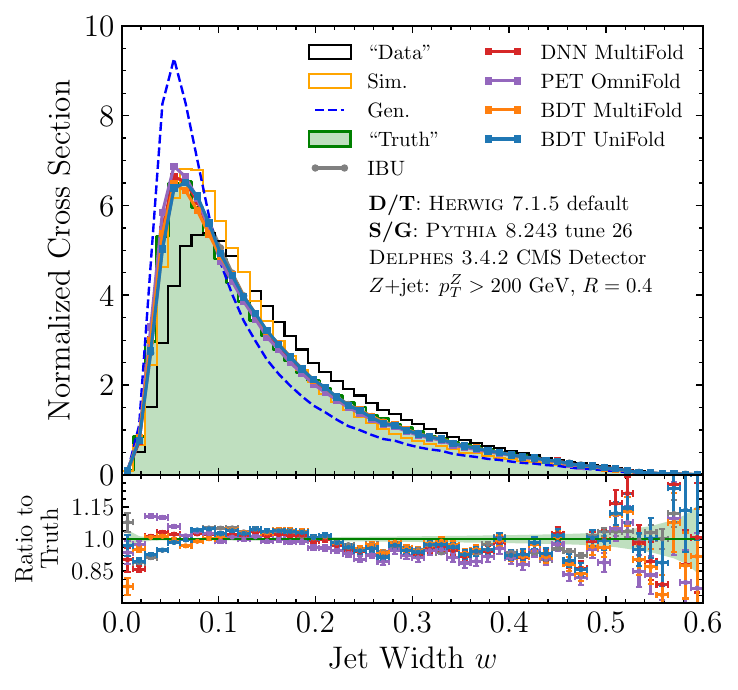}
    \end{minipage}\hfill
    \begin{minipage}{0.5\textwidth}
        \centering
        \includegraphics[width=\linewidth]{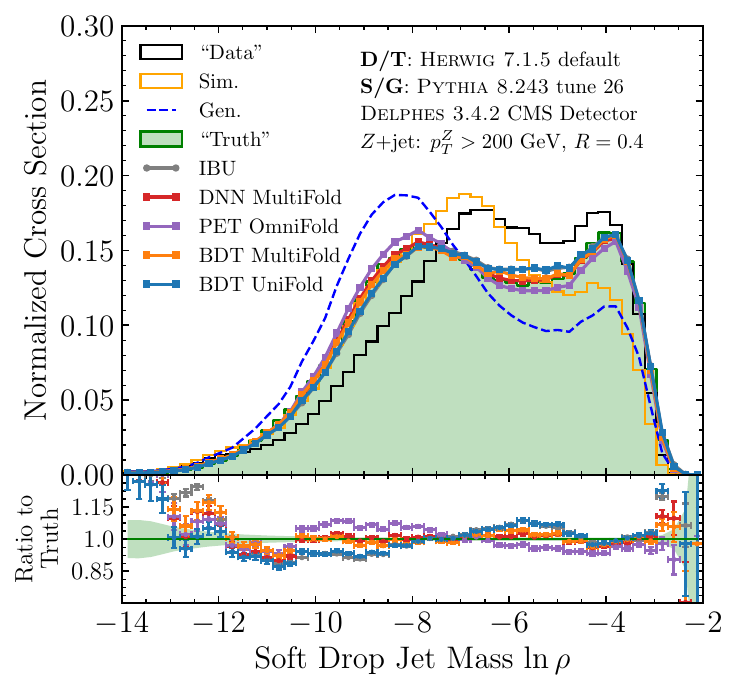}
    \end{minipage}

    \begin{minipage}{0.5\textwidth}
        \centering
        \includegraphics[width=\linewidth]{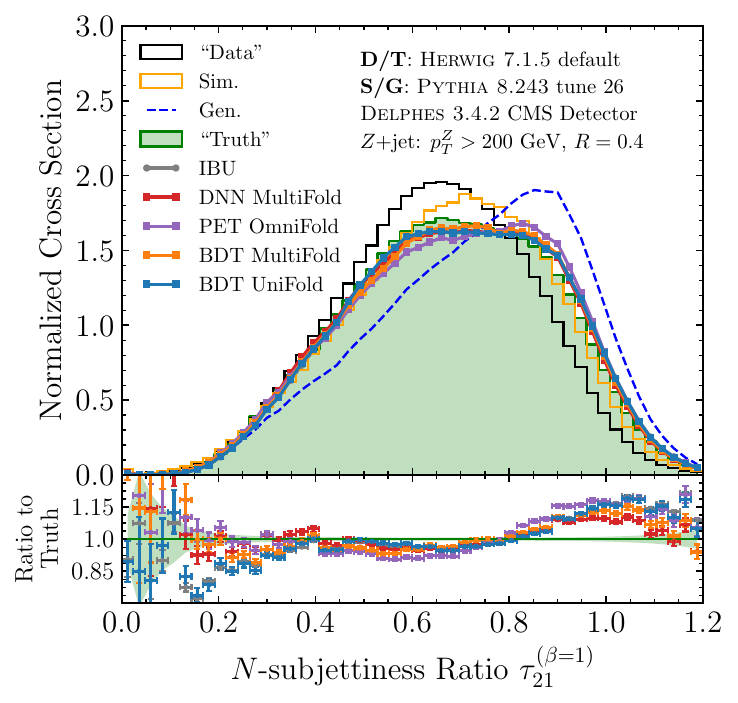}
    \end{minipage}\hfill
    \begin{minipage}{0.5\textwidth}
        \centering
        \includegraphics[width=\linewidth]{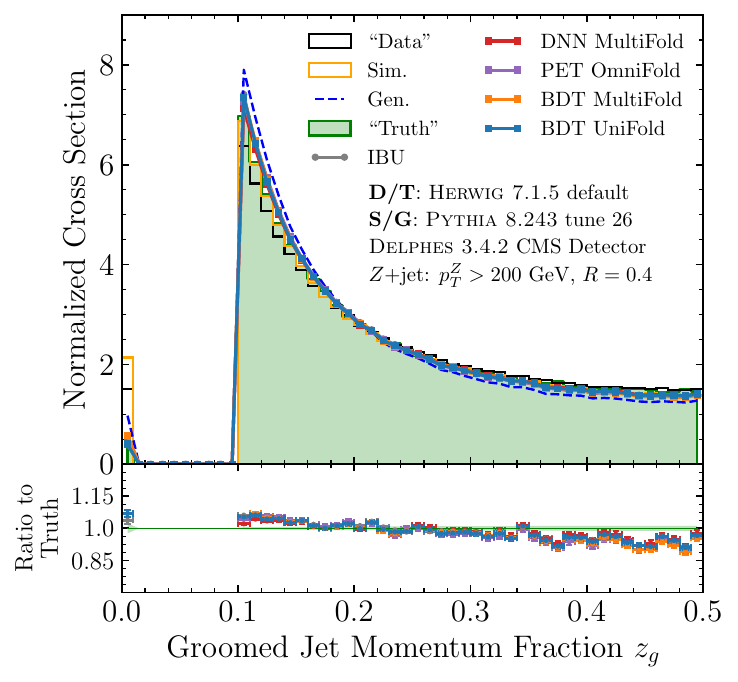}
    \end{minipage}

    \caption{Comparison of unfolded jet observables. \textsc{herwig} is used as the ``Data"/``Truth", while \textsc{pythia} is the simulation and generation (Sim./Gen.). The events shown here are from the test datasets, which contain about $400\mathrm{k}$ events.  The ratio panel only contains statistical uncertainties, which are described in Figure~\ref{fig:unfolded_gaussian}.}
    \label{fig:unfolded_observables}

\end{figure}

We compare the results of the unfolding in Fig.~\ref{fig:unfolded_observables}.  All of the methods show visibly good agreement with one another and are within about $\pm15\%$ of the truth values. As expected from the original OmniFold paper~\cite{Andreassen:2019cjw}, \textsc{OmniFold} and \textsc{MultiFold} generally show improved performance compared to the IBU and \textsc{UniFold} methods. The BDT and DNN accuracy for \textsc{MultiFold} are similar, as are the accuracy of the BDT \textsc{UniFold} and IBU.  These comparisons are quantified using the triangular discriminator~\cite{850703} in Table~\ref{table:metric}.  This metric is defined by $\frac{1}{2}\sum_{i} (p_i-q_i)^2/(p_i+q_i)$ (and multiplied by 1000), where $p_i$ and $q_i$ are the values of the binned truth distribution and the binned unfolded distribution in bin $i$, respectively.  As with the Gaussian data, the unbinned results here are binned for illustration purposes.  The \textsc{UniFold} result is unbinned per observable, the \textsc{MultiFold} result is one unfolding for all six observables, and the \textsc{OmniFold} result is one unfolding for the six observables and any other observable that can be computed for the phase space of all hadrons within the jets.

\section{Conclusion and Outlook}
\label{sec:conclusion}

In this paper, we have extended the toolkit of unbinned unfolding methods.  Our main contribution is to describe two pieces of software that serve as the official repositories of the OmniFold method: one based on ROOT and the RooUnfold package, which will be incorporated into RooUnfold, and one available in PyPI without dependence on ROOT. Along the way, we have studied decision trees as an alternative to neural networks for relatively low-dimensional cases.  We find comparable or even superior performance of BDTs compared to deep neural networks in terms of accuracy and strictly better performance in terms of computational cost.  To ensure our comparisons include all of the relevant aspects of unfolding, we have extended the standard $Z$+jets dataset to include acceptance effects, which all of our software implementations are able to accommodate.  Our goal is that these innovations will facilitate many unbinned unfolding measurements in the future and will also catalyze further methodological advancements.

\section*{Code and Data availability}
The BDT Omnifold class is available at \url{https://github.com/rymilton/unbinned_unfolding}. The code to train the DNN and PET models and to generate the plots in this paper are also in this repository. The neutral network versions of Omnifold are available in PyPI (\url{https://pypi.org/project/omnifold/}). The jet observable data are available on Zenodo~\cite{Zenodo:EnergyFlow:ZJetsDelphes}. 

\section*{Acknowledgments}

We thank Lydia Brenner and Vincent Croft for helpful discussions on RooUnfold and developing the BDT approach.  We also thank Fernando Torales Acosta and Jingjing Pan for useful discussions.  This material is based upon work supported by the National Science Foundation under Grant No. 2311666.

\FloatBarrier
\bibliography{HEPML.bib,other.bib}
\appendix
\section{Comparison of binned unfolding with BDT and IBU}
\label{sec:binned_unfolding}

While one main focus of the ML-based unfolding approaches is to avoid binning, we show its application to binned data in this appendix.  Our goal is two-fold: (1) to demonstrate that the method can accommodate binned inputs (indeed, you can supply histograms in the RooUnfold-inspired OmniFold and it will just operate on the discretized data) and (2) to show that OmniFold reduces to IBU in the binned limit.

We illustrate this using the BDT approach. To do this, we convert the response matrix (in the form of a 2D histogram) into lists of particle-level and detector-level simulation bin centers. We likewise convert the detector-level pseudodata histogram into a list of pseudodata bin centers. These lists contain the bin centers repeated $N$ times, where $N$ is the number of counts in the bin. This procedure effectively translates the binned unfolding into unbinned unfolding with the bin centers acting as the unbinned values. The normal \textsc{UniFold} procedure is carried out, and at the end of the unfolding we bin the lists with their unfolded weights and then divide each bin count by the efficiency vector obtained from the response matrix. This final step accounts for particle-level simulation events that were not reconstructed. 

\begin{figure}[h!]

    \centering
     \includegraphics[width=0.6\textwidth]{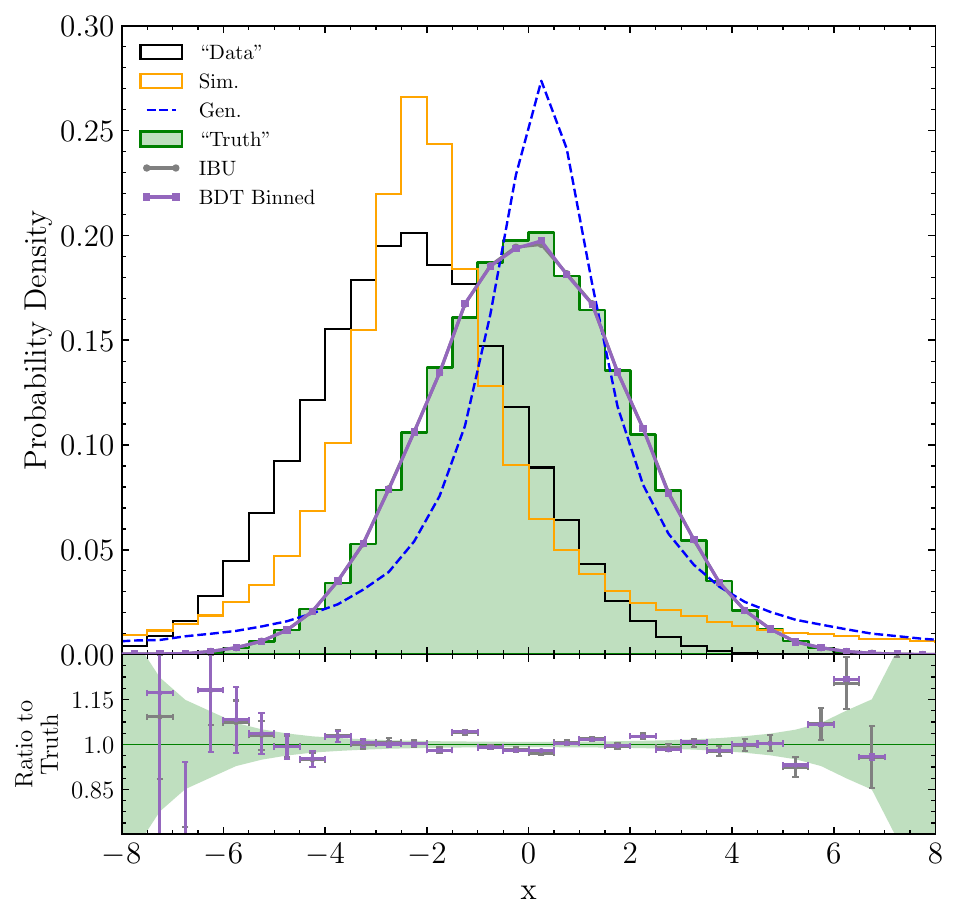}
     
     \caption{Comparison of unfolded Gaussians. Top panel: The distributions of the used datasets and the unfolded distributions using IBU and binned BDT. Bottom panel: Ratio of the unfolded distribution to the truth data in each bin. The statistical uncertainties are shown, as described in Figure~\ref{fig:unfolded_gaussian}.}
     \label{fig:gaussian_binned_unfolding}
\end{figure}

\begin{figure}[h]
    \centering
    \begin{minipage}{0.5\textwidth}
        \centering
        \includegraphics[width=\linewidth]{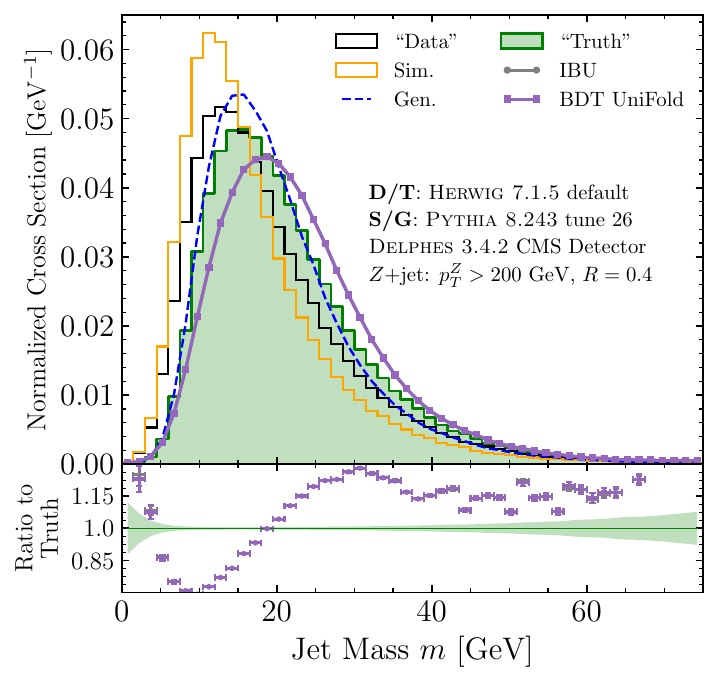}
    \end{minipage}\hfill
    \begin{minipage}{0.5\textwidth}
        \centering
        \includegraphics[width=\linewidth]{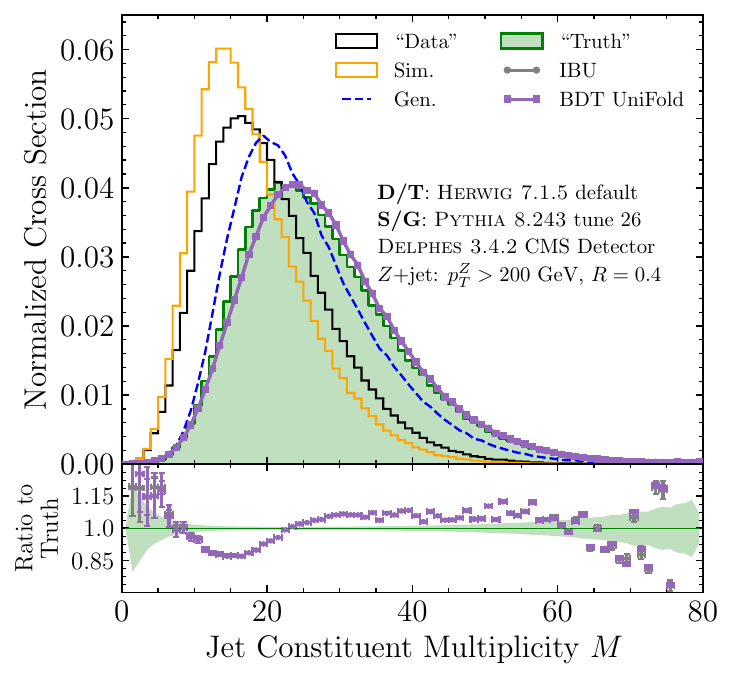}
    \end{minipage}

    \begin{minipage}{0.5\textwidth}
        \centering
        \includegraphics[width=\linewidth]{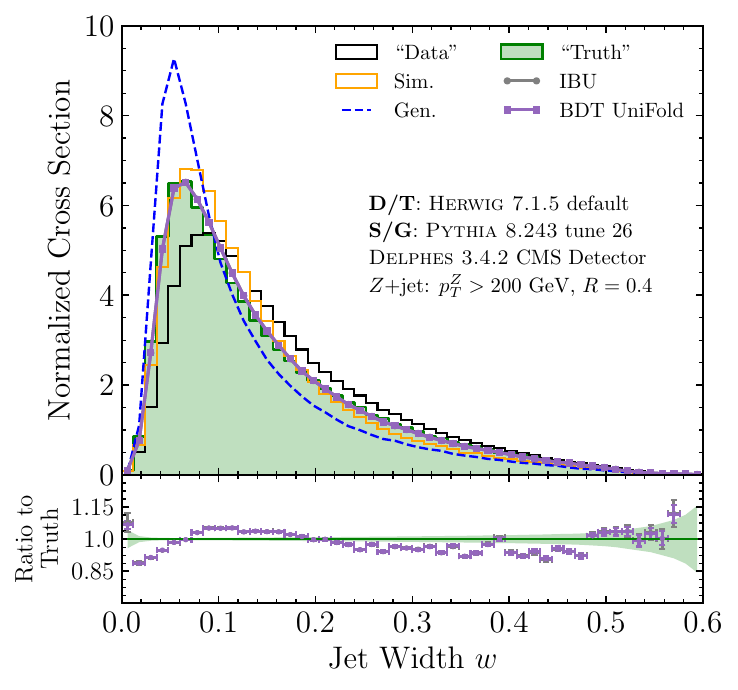}
    \end{minipage}\hfill
    \begin{minipage}{0.5\textwidth}
        \centering
        \includegraphics[width=\linewidth]{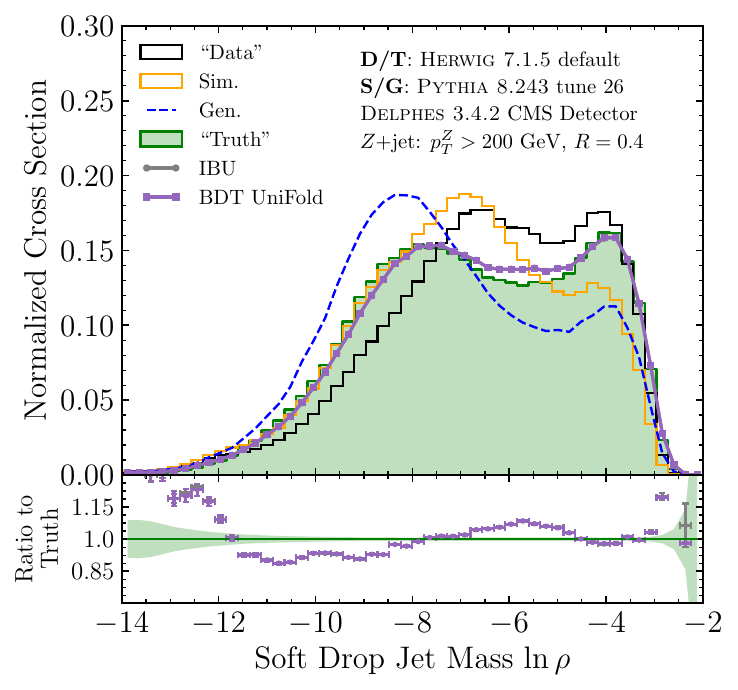}
    \end{minipage}

    \begin{minipage}{0.5\textwidth}
        \centering
        \includegraphics[width=\linewidth]{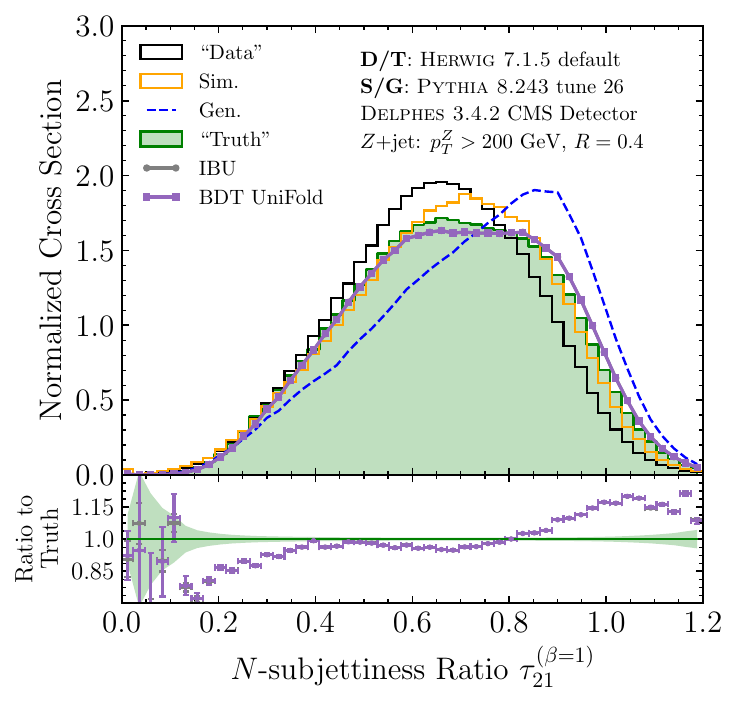}
    \end{minipage}\hfill
    \begin{minipage}{0.5\textwidth}
        \centering
        \includegraphics[width=\linewidth]{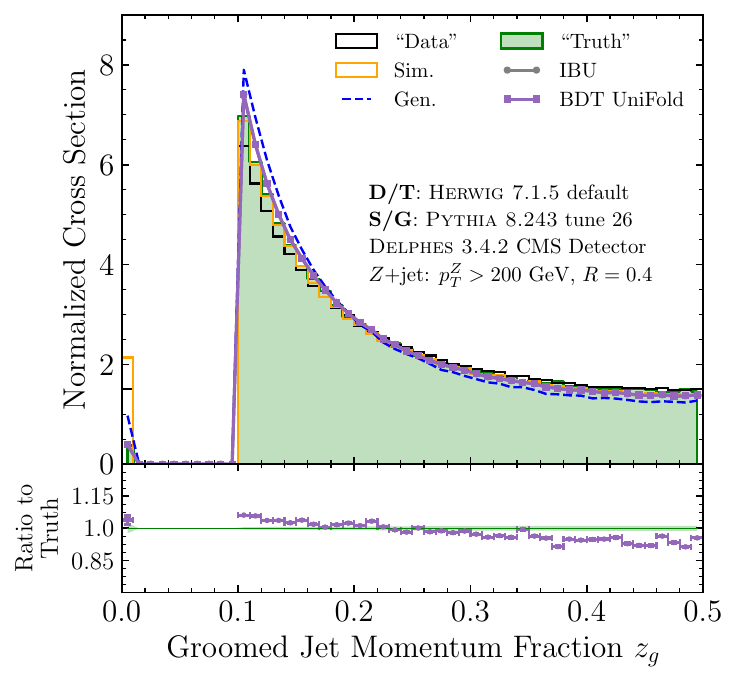}
    \end{minipage}

    \caption{Comparison of unfolded jet observables using the binned BDT method and IBU. For more details about the simulation data and pseudodata, see Fig.~\ref{fig:unfolded_observables}. The performance between the two unfolding methods is similar.}
    \label{fig:binned_unfolding_observables}

\end{figure}

We compare our binned unfolding with IBU using both the Gaussian distributions and the jet observables. The results are shown in Fig.~\ref{fig:gaussian_binned_unfolding} and Fig.~\ref{fig:binned_unfolding_observables}, respectively. In Table~\ref{table:binned_metric}, we quantify the performance of the methods with the triangular discriminator like before. As expected, binned \textsc{UniFold} reproduces IBU.

\begin{table}[h]
\centering
\footnotesize
\caption{Triangular discriminator comparison of binned unfolding methods. The metric is the same as described in Table~\ref{table:metric}. We also show the metric for the detector-level pseudodata and particle-level simulation data.}

\begin{tabular}{c c c c c c c}
     \hline
     \hline
     & \multicolumn{6}{c}{Observable} \\
     \cline{2-7}
     Method & $m$ & $M$ & $w$ & $\ln \rho$ & $\tau_{21}$ & $z_g$ \\
     \hline
      Binned BDT & 9.69 & 1.43 & 0.47 & 0.87 & 1.16 & 0.45 \\
    IBU & 9.74 & 1.46 & 0.48 & 0.89 & 1.15 & 0.45 \\
    \hline
    Data & 24.59 & 129.92 & 15.87 & 14.70 & 11.12 & 3.86 \\
    Generation & 3.46 & 14.94 & 22.39 & 18.97 & 20.70 & 3.99 \\
     \hline
     \hline
\end{tabular}
\label{table:binned_metric}
\end{table}

\section{Python examples of BDTs and Deep Neural Networks}
\label{sec:examples}
To demonstrate the usage of the RooUnfold-inspired BDT OmniFold class --- \newline \verb|RooUnfoldOmnifold| --- we provide examples of code for both binned unfolding (as done in Section~\ref{sec:binned_unfolding}) and unbinned unfolding. For simplicity, the code will be shown in Python using PyROOT, but it is a simple translation to regular ROOT.

We also show example code for unbinned unfolding with the pip version of OmniFold.

\subsection{BDT binned unfolding example}
Following the filling of the response matrix (\verb|response|) and a histogram containing the detector-level data (\verb|hMeas|), binned unfolding can be used as follows:
\begin{verbatim}
    num_iterations = 5
    omnifold = ROOT.RooUnfoldOmnifold(response, hMeas, num_iterations)
    hUnfolded = omnifold.Hunfold()
\end{verbatim}
\verb|hUnfolded| is a TH1 object with the unfolded data. This is identical to how other RooUnfold algorithms are used, allowing users to easily use the binned unfolding in \newline \verb|RooUnfoldOmnifold|.

\subsection{BDT unbinned unfolding example}
Unbinned unfolding utilizes lists rather than histograms as its inputs. Since each event can be multi-dimensional (e.g. simultaneously unfolding multiple observables), we use \verb|RDataFrames| for the input to the unbinned unfolding function. For quantities that are one dimensional, such as masks detailing reconstruction cuts or unfolded weights, we use \verb|TVectors|.

To configure the inputs, the user should make \verb|RDataFrames| for the particle-level simulation data (\verb|df_MCgen| below), detector-level simulation data (\verb|df_MCreco| below), and the detector-level measured data/pseudodata (\verb|df_measured| below). If needed, the user should also set up \verb|TVectors| to serve as masks for event cuts. For instance, if there are events in the simulation that do not satisfy detector-level cuts, these events should have a \verb|False| value in the mask \verb|TVector| (\verb|MC_pass_reco| below). The code below also shows cuts for the measured data (\verb|measured_pass_reco|) and particle-level simulation cuts (\verb|MC_pass_truth|). This latter cut simply masks out events that do not pass the particle-level cuts.
Note that the \verb|RDataFrames| for the truth-level and detector-level simulation data should have the same number of entries. Placeholder values for an event that does not pass particle-level or detector-level cuts can be used if no values exist.
\begin{verbatim}
    omnifold = ROOT.RooUnfoldOmnifold()
    omnifold.SetMCgenDataFrame(df_MCgen)
    omnifold.SetMCrecoDataFrame(df_MCreco)
    omnifold.SetMeasuredDataFrame(df_measured)
    omnifold.SetMCPassReco(MC_pass_reco)
    omnifold.SetMCPassTruth(MC_pass_truth)
    omnifold.SetMeasuredPassReco(measured_pass_reco)
    omnifold.SetNumIterations(5)
    unbinned_results = omnifold.UnbinnedOmnifold()
    step2_weights = ROOT.std.get[1](unbinned_results)
\end{verbatim}

The user can then fill a histogram with the particle-level simulation data with the weights from \verb|step2_weights| to get the unfolded distribution. 

By default, the trained BDTs will be saved to a pickle file in the present working directory. This path can be adjusted with the function 
\begin{verbatim}
    RooUnfoldOmnifold.SetSaveDirectory(directory_name)
\end{verbatim}
The models can then be used to generate weights for other data (i.e. test data), as seen below:

\begin{verbatim}
    omnifold.SetTestMCgenDataFrame(df_MCgen_test)
    omnifold.SetTestMCrecoDataFrame(df_MCreco_test)
    omnifold.SetTestMCPassReco(MC_test_pass_reco)
    test_unbinned_results = omnifold.TestUnbinnedOmnifold()
    step2_weights_test = ROOT.std.get[1](test_unbinned_results)
\end{verbatim}

Although the default parameters for the GradientBoostingClassifiers and GradientBoostingRegressors trees performed well for the studies in this paper, users can also adjust these parameters using a \verb|TMap|. In the example below, the \verb|n_estimators| parameter for the step 1 classifier is adjusted. The adjustable parameters for the GradientBoostingClassifier and GradientBoostingRegressor are the same parameters as in the associated scikit-learn classes. This also allows users to utilize validation data during training.

\begin{verbatim}
    step1classifier_params = ROOT.TMap()
    step1classifier_params.Add(
        ROOT.TObjString("n_estimators"), 
        ROOT.TObjString("50")
    )
    
    step2classifier_params = ROOT.TMap()
    step2classifier_params.Add(
        ROOT.TObjString("n_estimators"),
        ROOT.TObjString("50")
    )
    
    step1regressor_params = ROOT.TMap()
    step1regressor_params.Add(
        ROOT.TObjString("n_estimators"),
        ROOT.TObjString("50")
    )
    omnifold.SetStep1ClassifierParameters(step1classifier_params)
    omnifold.SetStep2ClassifierParameters(step2classifier_params)
    omnifold.SetStep1RegressorParameters(step1regressor_params)
\end{verbatim}

Lastly, initial weights for the datasets can be set with \verb|TVectorDs|:
\begin{verbatim}
    omnifold.SetMCgenWeights(initial_MCgen_weights)
    omnifold.SetMCrecoWeights(initial_MCreco_weights)
    omnifold.SetMeasuredWeights(initial_measured_weights)
\end{verbatim}

\subsection{PyPI unbinned unfolding example}
The pip version of OmniFold follows a similar structure as the RooUnfold class. We provide a dataloader object that takes as inputs numpy arrays containing the reconstruction level and generator level (when available) data. These objects can be created with few lines of code:

\begin{verbatim}
    from omnifold import DataLoader, MultiFold, MLP, SetStyle, HistRoutine
    # Assume the same inputs as before but in lists instead of ROOT objects: 
    # df_MCgen, df_MCreco, df_measured, 
    # initial_MCgen_weights, initial_measured_weights
    
    measured_dataloader = DataLoader(reco = df_measured, 
                                     weight = initial_measured_weights, 
                                     normalize=True)
    MC_dataloader = DataLoader(reco = df_MCreco, 
                                gen = df_MCgen, 
                                weight = initial_MCgen_weights, 
                                normalize=True)
\end{verbatim}
The $normalize$ flag is used to normalize the sum of weights to a common value between the two datasets, useful when measuring normalized differential cross-sections where the overall normalization is not used. This flag can be omitted for measurements where the overall normalization is also important. The dataloader also supports masking of the input entries to identify events not passing the event selection either at reconstruction or generation level named \verb|pass_reco| and \verb|pass_gen|, respectively. As the name suggests, values of 1 are set to events passing the selection while 0 is used for events not passing. The second step is to define the neural network architecture used for the training. We provide with the framework a simple MLP model as well as the PET model, but the user is free to create their own model as a \verb|keras.Model| object. Instantiating the model can be done as:

\begin{verbatim}
    model_reco = MLP(ndim_reco,
                    layer_sizes = [64,128,64],
                    activation = 'relu')
    model_gen = MLP(ndim_gen,
                    layer_sizes = [64,128,64],
                    activation = 'relu')
\end{verbatim}

where the model expects the number of features at reco and gen levels (\verb|ndim_reco| and \verb|ndim_gen|) as well as the dimensions of each hidden layer and the activation function applied at the end of each hidden layer. The next step is to create the \textsc{MultiFold} object:

\begin{verbatim}
    multifold = MultiFold("Run_name_for_your_experiment",
                         model_reco,
                         model_gen,
                         measured_dataloader,
                         MC_dataloader,
                         batch_size = 512,
                         lr = 5e-5,
                         niter = 5,
                         epochs = 100,
                         weights_folder = 'path_to_store_the_checkpoints')
\end{verbatim}

We can finally start the training using:

\begin{verbatim}
    multifold.Unfold()
\end{verbatim}
After the training, we can evaluate the last iteration of OmniFold to reweight a new set of simulations (or conversely, load any step from the saved checkpoints during step 2) using:

\begin{verbatim}
    unfolded_weights = multifold.reweight(validation_data,omnifold.model_gen)   
\end{verbatim}

The unfolded weights can then be used when plotting events from \verb|validation_data| to display the unfolded measurement. 
\end{document}